\documentclass[referee]{raa}
\usepackage{times,multirow}
\usepackage{natbib}
\usepackage{epstopdf,cite,mathtools,ifthen,xcolor,setspace}
\usepackage{textcomp,array,booktabs,epsfig,rotating,subfigure}
\usepackage{SIunits,color}
\usepackage{graphicx,epstopdf}
\bibpunct{(}{)}{;}{a}{}{,}

\begin{document}

\begin{spacing}{1}

\title{The Wide-field Photometric System of the Nanshan One-meter Telescope$^*$$^\dag$
\footnotetext{\small $*$ Supported by Key Laboratory of Optical Astronomy,
National Astronomical Observatories, Chinese Academy of Sciences.}
\footnotetext{\small $\dag$ Supported by the National Natural Science Foundation of China No. 11803076, 11873081 and U1831209.}}

 \volnopage{ {\bf 2018} Vol.\ {\bf X} No. {\bf XX}, 000--000}
   \setcounter{page}{1}

   \author{Chunhai Bai\inst{1,2}, Guojie Feng\inst{1}, Xuan Zhang\inst{1}, Hubiao Niu\inst{1}, Abdusamatjan Eskandar\inst{1}, Guangxin Pu\inst{1}, Shuguo Ma\inst{1}, Jinzhong Liu\inst{1}, Xiaojun Jiang\inst{2}, Lu Ma\inst{1}, Ali Esamdin\inst{1}, Na Wang\inst{1}
   }

   \institute{Xinjiang Astronomical Observatories,  Chinese Academy of Sciences,
150 Scientific 1st street, Urumqi, Xinjiang 830011, China; {\it baichunhai@xao.ac.cn, fengguojie@xao.ac.cn}\\
    \and Key Laboratory of Optical Astronomy, National Astronomical Observatories, Chinese Academy of Sciences, Beijing 100101, China\\
\vs \no
   {\small Received 2020 Mar. 3; accepted 2020 May 27}
}
\abstract{The Nanshan One-meter Wide-field Telescope (NOWT) is a prime focus system located at Nanshan Station of Xinjiang Astronomical Observatories (XAO). The field of view(FOV) was designed to 1.5$\degree$ $\times$ 1.5$\degree$, and Johnson-Cousins UBVRI system was chosen as the main Filter set. The telescope has been providing observation services for astronomers since Sept. 2013. Variable source searching and time-domain surveys are the main scientific goals. The system's test results are reported including linearity, dark current, bias, readout noise and gain of the CCD camera. The accurate instrumental calibration coefficients in UBVRI bands was driven with Landolt standard stars during photometric nights. Finally, the limiting magnitudes are given with signal-to-noise ratios and various exposure times for observers.
\keywords{telescope; filter; CCD photometry.}
}
   \authorrunning{C.-H. Bai et al.}            
   \titlerunning{Photometric System on Nanshan One-meter Wide-field Telescope}  
   \maketitle
%

\section{INTRODUCTION}           
\label{sect:intro}

The NOWT situated at Nanshan station (87$\degree$10$\arcminute$30$\arcsecond$ East, 43$\degree$28$\arcminute$25$\arcsecond$ North; at an altitude of about 2088m) of XAO (\citealt{Hu:2017ART}, \citealt{Liu:2014IAUS}), Chinese Academy of Sciences (CAS). The distribution peak of seeing values around 1.67\arcsecond, and 80\% of the value obtained at nights are below 2.2\arcsecond. At zenith, the sky brightness  is around 21.7 magarcsec$^{-2}$ in the V-band.  The observable night of Nanshan Station is more than 300 days per year, and the clear night is greater than 210, according to the observation records of a 40-cm optical telescope (2005 $\sim$ 2008). Compared to other domestic one meter class telescopes, NOWT has a larger field of view and a higher coverage efficiency in sky survey. At present, the telescope has a mature observation and technical maintenance team. According to the needs of different data reduction, the corresponding pipeline has been completed. The main scientific projects conducted using NOWT after the installation have been multi-color photometry of binaries (\citealt{Zhang:2018RAA}), pulsating stars (\citealt{Fu:2017ASPC}), exoplanet (\citealt{Wang:2017}), variable stars (\citealt{Yang:2018RAA}, \citealt{Bai2019}), gravitational wave ( \citealt{Liu:2014PIAU}), galaxies(\citealt{Zhang:2015MNRAS}), survey(\citealt{Ma:2018}, \citealt{Zheng2019RAA}).

In Mar. 2012, the telescope was installed by Germany APM telescope company and XAO optical technicians. It would be useful and significant if observers could be aware of the performance and characteristics of NOWT, including the dark, bias level and linearity of the camera. Besides, the CCD photometric system was analyzed carefully and results are reported, including the instrument response, throughput and detection limit.

In Section \ref{sect:Obs}, the NOWT observation system was introduced. In Section \ref{sect:CCD}, the specifications of camera are reported. The transformation coefficients and calibration of photometrical system are given in Section \ref{sect:PHOTOMETRIC}. In Section \ref{sect:PERFORMANCE}, we gave the detection limit and throughput. In Section \ref{sect:SUMMARY}, a summary is given.

\section{PRIME-FOCUS OPTICS AND CAMERA}
\label{sect:Obs}

\begin{figure}
    \begin{minipage}[t]{0.45\linewidth}
    \begin{center}
        \scalebox{1}[1]{\includegraphics*[bb=0 0 526 760,width=1\textwidth,clip]{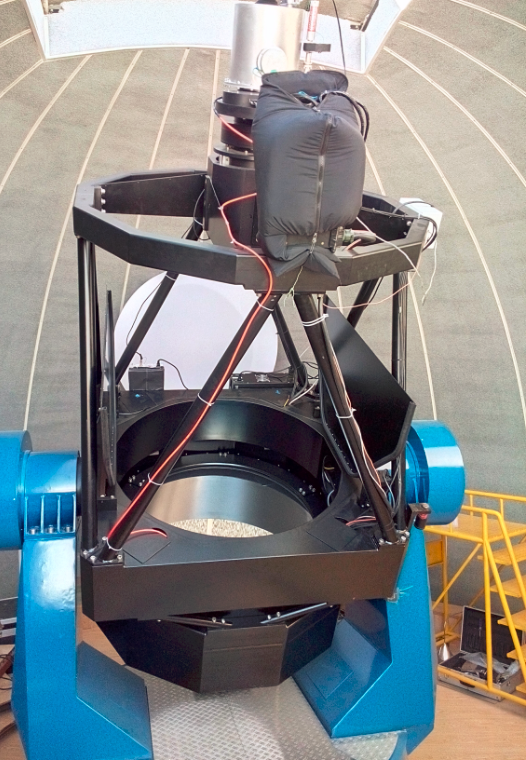}}
        \caption[Fig. Bias1]{{\small Front view of NOWT.}}
    \end{center}
    \label{fig:Nowt_front}   
    \end{minipage}
    \begin{minipage}[t]{0.55\linewidth}
    \begin{center}
        \scalebox{0.9}[0.9]{\includegraphics*[bb=0 0 385 393,width=1\textwidth,clip]{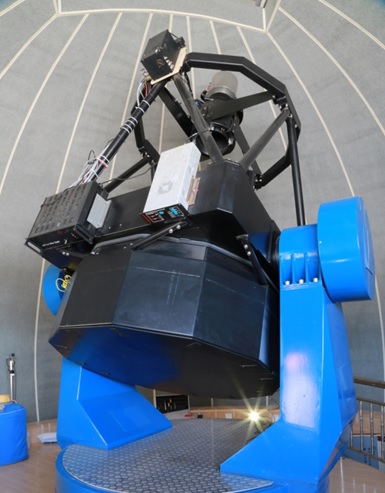}}
        \caption[Fig. Bias2]{{\small Back view of NOWT.}}
    \end{center}
    \label{fig:Nowt_Back}  
    \end{minipage}
\end{figure}

The NOWT is a horizontal mounting telescope (Fig. 1 and \ref{fig:Nowt_Back}). The parabolic primary mirror's effective diameter is 1000mm, 80\% of the collected energy is concentrating into a circle with diameter of less then 1.15$\arcsecond$ on a field diameter of 2.4$\degree$. The prime focus designed focal ratio is 2.2, but the actually focal length of NOWT is 2159 $\pm$ 20mm. The mirrors and mount characteristics of NOWT are listed in Table 1. The CCD camera was designed and integrated by the CCD laboratory of the National Astronomical Observatories of China(NAOC), CAS. The CCD chip model is E2V CCD203-82, and is blue sensitive, with 4 amplifiers and 16 bit A/D converter. The spectral response of the chip is shown in Figure 3.  The CCD is a scientific-grade chip, but don't have anti-blooming gate. It was made by E2V technologies company, and mounted to NOWT in Oct. of 2012. The specifications of CCD camera are listed in Table \ref{table:CCDmanual}. It was cooling by liquid nitrogen, the temperature can reach -110\textdegree C in summer and -135\textdegree C in winter. With a 4096 $\times$ 4136 imaging pixels (12 $\times$ 12 {$\mu$}m pixel$^{-1}$ ), the efficiency FOV of the CCD is 78$\arcminute$ $\times$ 78$\arcminute$, and the pixel scale is 1.125$\arcsecond$. A Johnson-Cousins standard UBVIR filters system that was made by Custom Scientific company started to serve since Jan. of 2013. Whenever needed, time synchronization is realized through networks which was built based on a local GPS time server.

\begin{center}
\begin{tabular}{lll}
\multicolumn{2}{c}{{\bf Table 1} Characteristics of the NOWT Mirrors and Mount}\\
\hline
Features & Characteristics \\
\hline
        effective diameter &  1000 mm \\
        prime focal length	& 2200 mm \\
        FOV at prime focus &	1.5$\degree$ $\times$ 1.5$\degree$ \\
        Efficiency of prime focus	& 87.46\% \\
        primary mirror material	& Schott Zerodur \\
        maximum slew speed & $>$7$\degree$ /s (both axes)   \\
        maximum accelerations Azimuth & $>$1.8$\degree$ /s$^{2}$ \\
        maximum accelerations Altitude & $>$1.3$\degree$ /s$^{2}$ \\
        pointing accuracy RMS-Error of RA & 2.4\arcsec \\
        pointing accuracy RMS-Error of DEC & 3\arcsec \\
        tracking accuracy RA & 1.8\arcsec RMS in 60 minutes \\
        tracking accuracy DEC & 1.8\arcsec RMS in 60 minutes \\
        zenith blind hole & $<$2$\degree$ \\
        Rotation angle on Azimuth &	+/- 289$\degree$\\
        Rotation angle on Elevation &	7.5$\degree$ $\sim$ 90.7$\degree$ \\
        rotater angle	& +/- 180$\degree$ \\
        maximum rotation speed &	2$\degree$/s \\
        focus accuracy & 1{$\mu$}m \\
\hline
\label{table:MountMirror}
\end{tabular}
\end{center}

\begin{center}
\begin{tabular}{lll}
\multicolumn{2}{c}{{\bf Table 2} Specifications of the Chip According to the Manual}\\
\hline
Features & Specifications \\
\hline
    Pixel number    & 4096 $\times$ 4136  \\
    Pixel size      & 12 {$\mu$}m $\times$ 12 {$\mu$}m  \\
    Pixel scale	& 1.125$\arcsecond$ \\
    Imaging area   & 49.2mm $\times$ 49.6mm   \\
    Effective FOV  & 78$\arcminute$ $\times$ 78$\arcminute$ \\
    Fill factor     & 100\%  \\
    A/D conversion  & 16 bit  \\
    Number of output amplifiers & 4 \\
    Number of serial registers  & 2 \\
    Number of serial underscan pixels & 50 \\
    Shortest exposure time      & 100ms     \\
    Full well (min $\sim$ typical)         & 130,000 $\sim$ 175,000 e$^-$ pixel$^{-1}$  \\
    Scan rates (Fast,Medium,Slow)                 & 146 kHz, 91 kHz, 51 kHz\\
    Full frame readout time     & 27s@146kHz, 44s@91kHz, 78s@51kHz \\
    Operating temperature       & -100$\celsius$ $\sim$ -130$\celsius$ \\
    Readout noise               & 3 $\sim$ 4.5e$^-$ \\
    Dark current (e$^-$/pixel/hr)    & 3 @ -100\celsius, 0.01 @ -120\celsius \\
    Linearity                   & $>$ 99.9995 \% \\
    Spectral range              & 300 $\sim$ 1060 nm \\
    Peak quantum efficiency     & 90 \% \\
\hline
\label{table:CCDmanual}
\end{tabular}
\end{center}

\begin{figure}
    \begin{center}
        \scalebox{0.8}[0.8]{\includegraphics*[bb=0 0 1299 663,width=1\textwidth,clip]{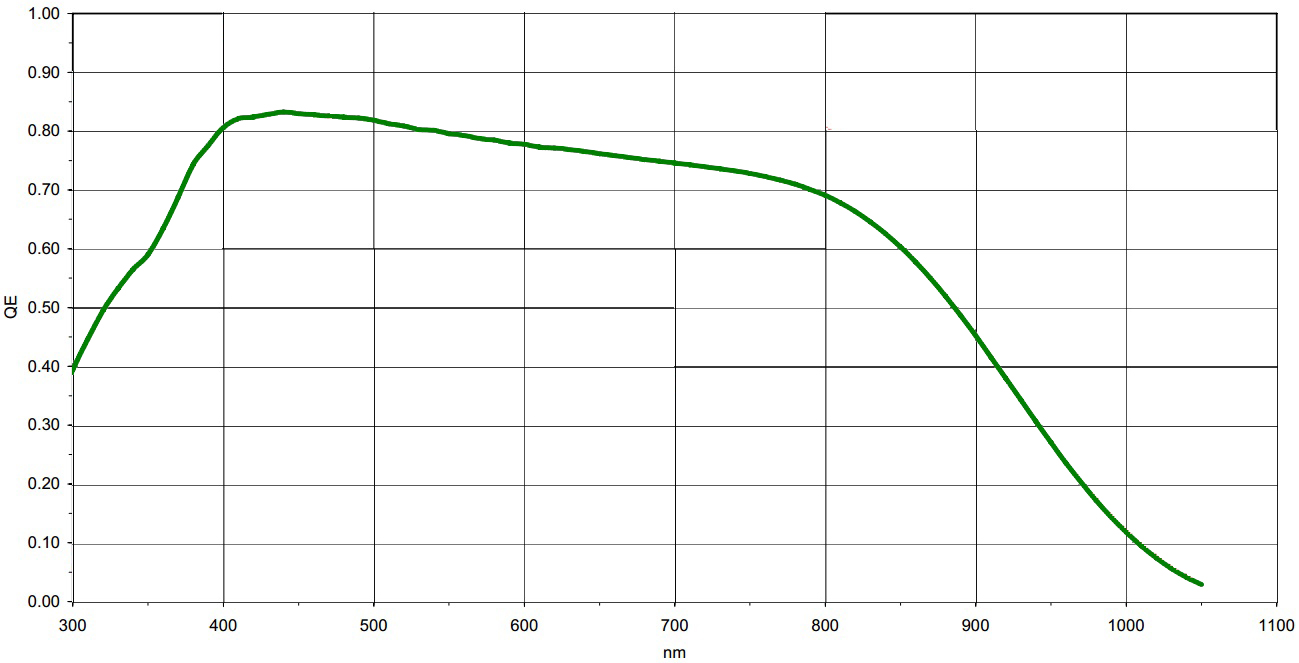}}
        \caption[Fig. QE]{{\small The spectral response of chip is shown as the solid green line.}}
    \end{center}
    \label{fig:Nowt_CCD_QE}  

\end{figure}

\section{CAMERA CHARACTERISTICS}
\label{sect:CCD}
\subsection{Bias}

We can get the original bias image with over scan in Figure 4. In order to make the over scan region more obvious, we used the 2k x 2k image. When the ambient temperature changes relatively large, the bias value also changes slightly. So we set 32 column under scan pixels for each amplifier, one can use the over scan to correct image data. Bias level of different reading speed and gain setting were tested on the night of 2017 June, we take ten bias images for each model. The bias images which corrected by overscan was shown in Figure 5. We can see that the proceeded image was acceptable.


\begin{figure}
   \begin{minipage}[t]{0.55\linewidth}
    \begin{center}
        \scalebox{1}[1]{\includegraphics*[bb=0 0 981 949,width=1\textwidth,clip]{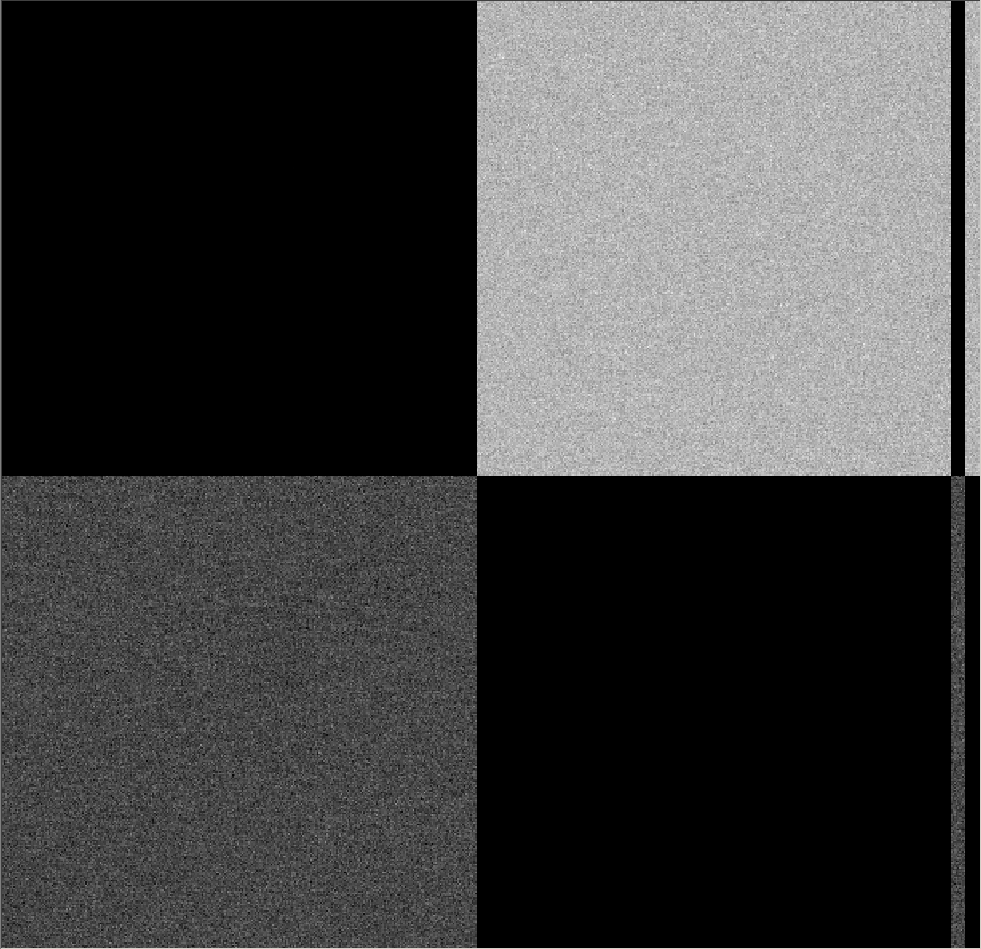}}
        \caption[Fig. Biasover]{{\small Bias image with over scan of four amplifiers.}}
    \end{center}
        \label{fig:BIAS_over}    
    \end{minipage}
    \begin{minipage}[t]{0.45\linewidth}
    \begin{center}
        \includegraphics[width=6cm]{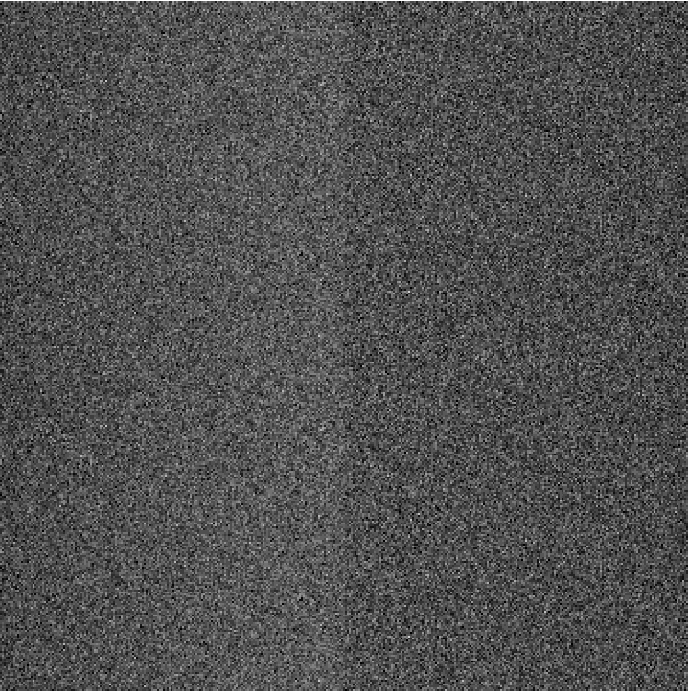}
        \caption[Fig. Bias2]{{\small Four amplifiers bias corrected by overscan, the ADU is uniform.}}
    \end{center}
    \label{fig:Bias-ov}  
    \end{minipage}
\end{figure}

The bias stable of the chip was tested in the evening of 2017 Sep. , when the temperature of CCD chip is lower than -100\celsius. As shown in Figure 6, the bias changed slightly when the temperature of CCD chip dropped from -117{\celsius} to -121{\celsius}.  Because the CCD has four gates, we have tested the stability of the four amplifiers. UL, UR, DL and DR represents the up left amplifier, the up right amplifier, the lower left amplifier and the lower right amplifier respectively. TDET represents the temperature of detector. One can see that the bias values of the four amplifiers are very stable on their respective Analog-to-Digital (ADU). The ADU of the four amplifiers is different because the gain of the four gates is slightly different, which will be described in detail in the below gain section. The scan rates and gain selection were set to Medium and 1x, because many observers use this selection.

\begin{figure}
    \begin{center}
        \includegraphics[width=12cm]{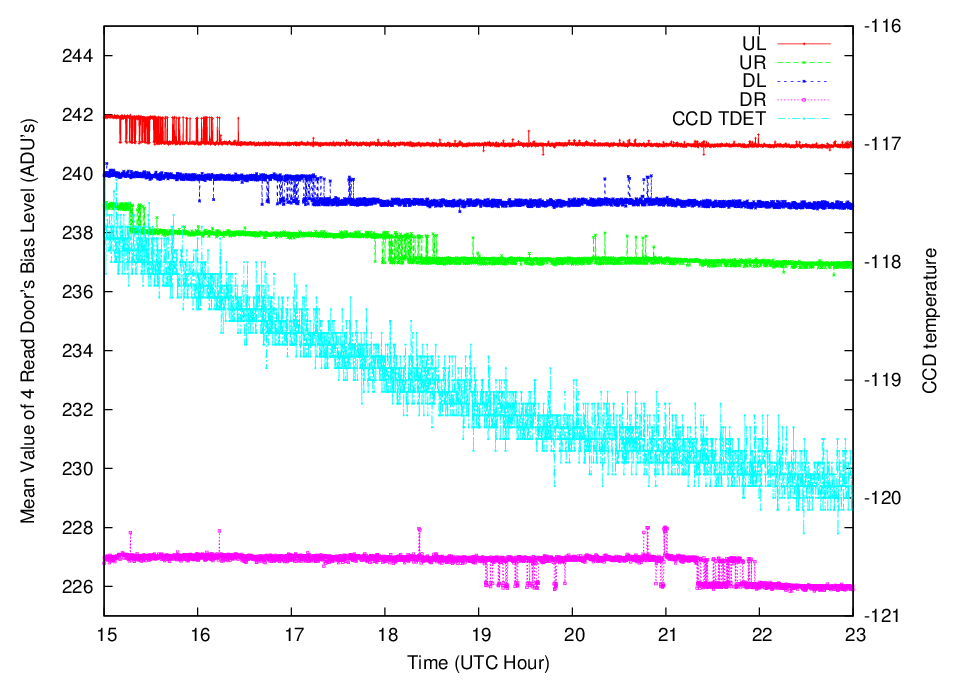}
        \caption[Fig. Bias3]{{\small Four amplifiers bias level when CCD temperature dropped from -117{\celsius} to -121{\celsius}, the blue curve shows the CCD temperature. Bias is stable during the observation.}}
    \end{center}
    \label{fig:Bias_TCCD}  
\end{figure}

\subsection{Readout Noise and Gain}
\label{sec:gainRN}

The gain(G) is useful for observer to evaluate the camera's performance, it is the factor that describes how many the digital outputs are converted into one unit electrons. The readout noise and gain values was calculated by Image Reduction and Analysis Facility's (IRAF's, provided by NOAO) task {\it findgain}, based on Equations \ref{equ:gain} and \ref{equ:RN} (\citealt{Howell:2000}, \citealt{Howell:2006}). The principle is to compare the signal levels and the amount of variation in the two bias and two flat field images.

The chip has four amplifiers, so for each reading speed and gain selection, we tested the gain exact value of each amplifier. The testing was carried out on night of 2017 June 26th. For the three reading modes of Fast, Medium and Slow, we have selected different gain options to obtain the actual gain value and readout noise. The results are listed in Table 3. It shows that all the readout noise value are similar with CCD chip nominal value, as given in Table \ref{table:CCDmanual}.

\begin{eqnarray}
Gain = \frac{(\overline{F1}+\overline{F2})-(\overline{B1}+\overline{B2})}{\sigma_{F1-F2}^{2} - \sigma_{B1-B2}^{2}}
\label{equ:gain}
\end{eqnarray}
\begin{eqnarray}
Read out noise = \frac{ Gain \cdot \sigma_{\overline{B1} - \overline{B2}}}{\sqrt {2}}
\label{equ:RN}
\end{eqnarray}

\begin{table}[!htbp]
\centering
\begin{tabular}{cccc}
\multicolumn{3}{c}{\bf Table 3} Test Results About Gain and Readout Noise\\
\hline
Models & Amplifiers & Gain(e$^-$/ADU) & Readout Noise(e$^-$) \\
\hline
\multirow{4}{*}{Fast (146Kpix/s G 0)} & UL & 0.980 & 3.030 \\
 & UR & 0.970 & 3.515 \\
 & DL & 0.990 & 4.345 \\
 & DR & 0.980 & 4.150 \\
 \hline
\multirow{4}{*}{Fast (146Kpix/s G 1)} & UL & 0.470 & 2.790 \\
 & UR & 0.460 & 3.440 \\
 & DL & 0.470 & 4.425 \\
 & DR & 0.460 & 4.220 \\
 \hline
\multirow{4}{*}{Fast (146Kpix/s G 2)} & UL & 0.210 & 2.740 \\
 & UR & 0.200 & 3.340 \\
 & DL & 0.210 & 4.290 \\
 & DR & 0.210 & 4.010 \\
 \hline
\multirow{4}{*}{Medium (91Kpix/s G 0)} & UL & 3.610 & 4.380 \\
 & UR & 3.585 & 4.430 \\
 & DL & 3.680 & 4.755 \\
 & DR & 3.640 & 4.840 \\
 \hline
\multirow{4}{*}{Medium (91Kpix/s G 1)} & UL & 1.675 & 2.885 \\
 & UR & 1.660 & 3.110 \\
 & DL & 1.715 & 3.375 \\
 & DR & 1.695 & 3.400 \\
 \hline
\multirow{4}{*}{Medium (91Kpix/s G 2)} & UL & 0.755 & 2.525 \\
 & UR & 0.755 & 2.640 \\
 & DL & 0.775 & 3.005 \\
 & DR & 0.770 & 3.010 \\
 \hline
\multirow{4}{*}{Slow (51Kpix/s G 0)} & UL & 1.590 & 2.610 \\
 & UR & 1.455 & 2.435 \\
 & DL & 1.620 & 2.770 \\
 & DR & 1.500 & 2.740 \\
 \hline
\multirow{4}{*}{Slow (51Kpix/s G 1)} & UL & 0.830 & 2.360 \\
 & UR & 0.800 & 2.335 \\
 & DL & 0.835 & 2.530 \\
 & DR & 0.810 & 2.660 \\
 \hline
\multirow{4}{*}{Slow (51Kpix/s G 2)} & UL & 0.390 & 2.270 \\
 & UR & 0.390 & 2.300 \\
 & DL & 0.400 & 2.430 \\
 & DR & 0.390 & 2.680 \\
\hline
\label{table:gain}
\end{tabular}
\end{table}

\subsection{Dark current}

On the night of 2018 June  and July , we took the dark current data for the whole night. The measured dark current of four amplifiers were listed in Table 4, the fitting image were listed in Figure 7. From the test data, dark current signal is so small that we can't get the dark generate rate when the integral time less than 1800s. We use the high gain to detect the dark current value, the CCD temperature varied between -100{\celsius} and -104{\celsius} during the test. The dark data were taken with the exposure time 3600s and 7200s. For each exposure time, 3 frames were taken and more bias for correction. Hence, we got the mean and standard deviations from each image. One can see that the dark current approach the product manual.

\begin{center}
\begin{tabular}{cccc}
\multicolumn{3}{c}{{\bf Table 4} The Dark current of four amplifiers for Slow Reading speed and Gain 2}\\
\hline
Amplifier & Dark current ADU pixel$^{-1}$ s$^{-1}$ & Dark current e$^-$ pixel$^{-1}$ hour$^{-1}$\\
\hline
UL & 0.000092 $\pm$ 0.0000276 & 0.1325 $\pm$ 0.03974\\
UR & 0.000135 $\pm$ 0.0000331 & 0.1944 $\pm$ 0.04766\\
DL & 0.000135 $\pm$ 0.0000344 & 0.1944 $\pm$ 0.04954\\
DR & 0.000163 $\pm$ 0.0000360 & 0.2347 $\pm$ 0.05198\\
\hline
\label{table:Dark_curt}
\end{tabular}
\end{center}

\begin{figure}[htbp]
    \small
    \centering
        \includegraphics[width=15cm]{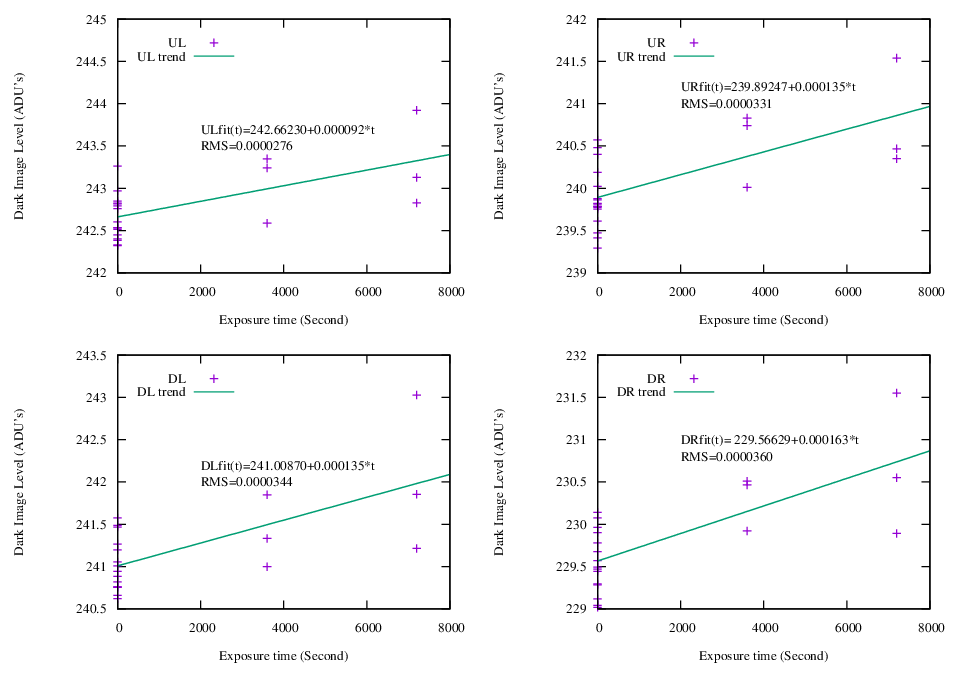}
        \caption[Fig. Darkfour]{{\small Dark current of four amplifiers for Slow Reading speed and Gain 2.}}
    \label{fig:Dark_four}   
\end{figure}

\subsection{Linearity}

For high precision photometry, CCD's linearity is very important. In order to accurately test the response's linearity of camera, measurements were carried out over two nights in June 2017 with dome flats. We took the measurements on night, because the stray light will change with the Sun moving in daytime. When dusk came, we first closed the cupola. Then we used A4 paper to adjust the brightness of the incandescent lamp in the dome, made the light homogeneous and dim enough for the CCD camera that could be take long time exposure. One tests is the linearity of the full well, the other test is the linearity of the A/D conversion.  The mean ADU values were counted based on a 400 x 400 region close to CCD center of each amplifier area.

For full well test, we set CCD linearity with medium read out speed and gain of 3.6.  The exposures was set from 0s to 90s. The mean count increased with the exposure time, after CCD pixels reached full well, the mean ADU value deviate the main fit line. In Figure 8, we can get that the linearity peak is larger than 36000, the peak value multiplied by 3.63 equals 130 635 ( $>$ 130 000 ). So, the full well linearity of the CCD camera is better than 99.999\%.

For linearity of the A/D conversion, we set CCD parameters with reading speed of Medium and Gain 1, which the gain values approached 1.69 according to the Table 3. The exposure time was set from 0s to 50s. After the AD converter saturated at 65535, the 16 bit ADU value decreases as shown in Figure 9. The electron count is close to 110000 and is far from the full well of 130000, but We understand that the A/D conversion reached it limited.

\begin{figure}
    \begin{minipage}[t]{0.5\linewidth}
        \includegraphics[width=7cm]{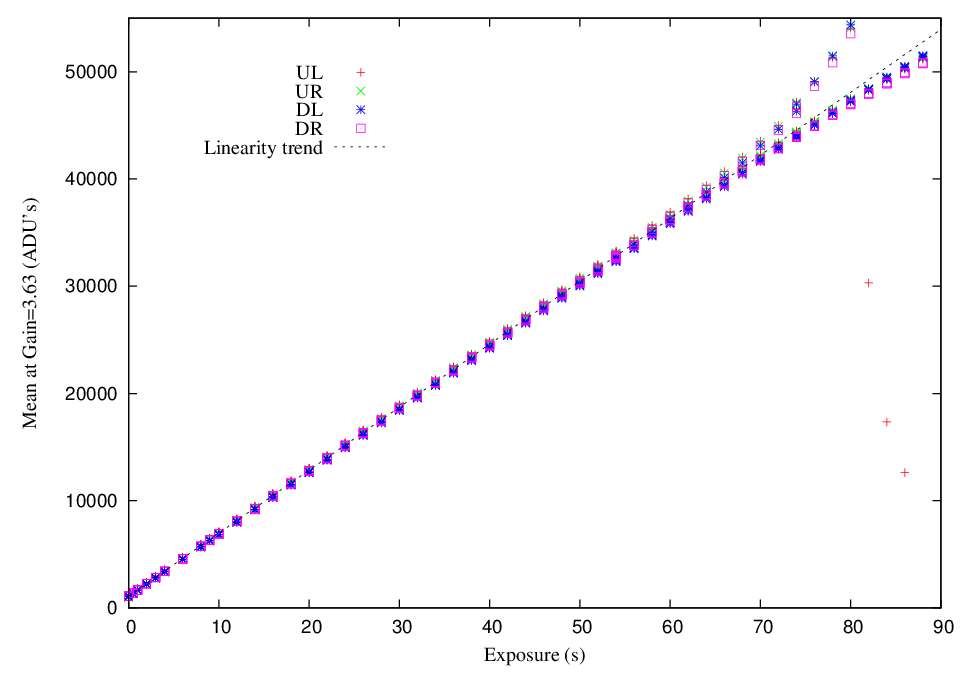}
        \caption[Fig. linearity3.4]{{\small Mean vs exposure with CCD reading speed of Medium with Gain 0. }}
    \label{fig:linearityG1}   
    \end{minipage}
    \begin{minipage}[t]{0.5\linewidth}
        \includegraphics[width=7cm]{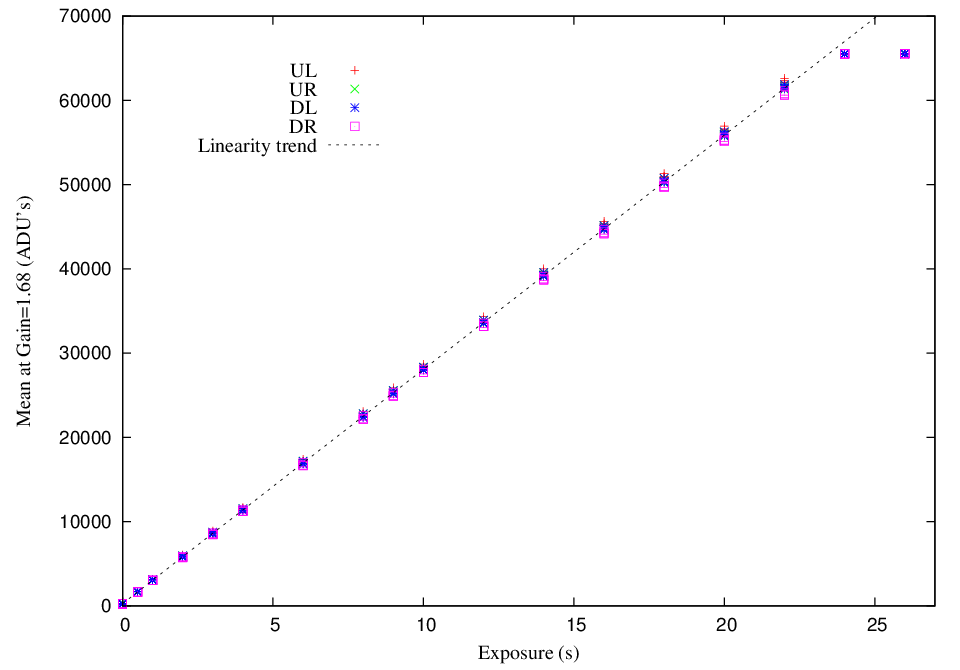}
        \caption[Fig. linearity0.97]{{\small Mean vs exposure with CCD reading speed of Medium with Gain 1. }}
    \end{minipage}
    \label{fig:linearityG4}
\end{figure}

\section{CALIBRATION OF PHOTOMETRY}
\label{sect:PHOTOMETRIC}

The magnitudes which we obtain directly from the images of the CCD cameras is commonly called as instrumental magnitudes. If observer plan to cross the NOWT result with history database or other telescope, he needs to know the calibrate coefficients of NOWT instrumental magnitudes to standard system. In May and Aug. 2017, we selected dozens of standard stars from Landolt catalog(\citealt{landolt:1992}, \citealt{landolt:2013}) which were listed in Table 5, considering both the magnitude and colors. Only some of the standards were listed in Table 5. For all the targets has been observed, please refer to the appendix. Some of standard stars can pass the Nanshan station's zenith, so they are very suitable for calibration.

\begin{center}
\begin{tabular}{lcccccccc}
\multicolumn{9}{c}{{\bf Table 5} Landolt Standards Used in 2017}\\
\hline
Star-Name & $\alpha$ (2000) & $\sigma$ (2000) & \it{V} & \it{B - V} & \it{U - B} & \it{V - R} & \it{R - I} & \it{V - I}\\
\hline
    GD 310C   & 11:29:25 & +38:06:12 & 13.911 & +0.892 & +0.668 & +0.496 & +0.426 & +0.924 \\
    GD 310A   & 11:29:29 & +38:09:14 & 14.057 & +0.540 & -0.073 & +0.341 & +0.356 & +0.698 \\
    GD 310B   & 11:29:35 & +38:08:12 & 14.200 & +0.978 & +0.832 & +0.595 & +0.541 & +1.137 \\
    SA 32-166 & 12:56:08 & +44:00:55 & 16.316 & +0.522 & -0.165 & +0.409 & +0.337 & +0.895 \\
    \ldots    &\ldots    & \ldots    & \ldots & \ldots & \ldots & \ldots &\ldots  & \ldots \\
\hline
\label{table:obsed3}
\end{tabular}
\end{center}

On observation night, the seeing values approximated 2.0\arcsecond. After collecting all the standard stars photometric data, we used the standard photometry steps in {\it apphot} package to reduce all images in IRAF, which the aperture was set 2 times of seeing. 25 is taken as zero point of instrumental magnitudes. Then we obtained the airmass, instrumental magnitudes and errors. The transformation equations was defined as follows,

\begin{eqnarray}
 U_{i}&=&U_{s}+Z_U+K^{'}_{U}X+C_U(U-B)_{s} \\
 \label{equ:U}
 B_{i}&=&B_{s}+Z_B+K^{'}_{B}X+C_B(B-V)_{s} \\
 \label{equ:B}
 V_{i}&=&V_{s}+Z_V+K^{'}_{V}X+C_V(B-V)_{s} \\
 \label{equ:V}
 R_{i}&=&R_{s}+Z_R+K^{'}_{R}X+C_R(V-R)_{s} \\
 \label{equ:R}
 I_{i}&=&I_{s}+Z_I+K^{'}_{I}X+C_I(V-I)_{s}
 \label{equ:I}
\end{eqnarray}

where U$_{s}$, B$_{s}$, V$_{s}$, R$_{s}$ and I$_{s}$ are the Landolt standard magnitudes, U$_{i}$, B$_{i}$, V$_{i}$, R$_{i}$ and I$_{i}$ are the NOWT instrumental magnitudes, Z$_U$, Z$_B$, Z$_V$, Z$_R$ and Z$_I$ are zero point magnitudes of NOWT, K$^{'}_{U}$, K$^{'}_{B}$, K$^{'}_{V}$, K$^{'}_{R}$ and K$^{'}_{I}$ are the first-order coefficients of extinction , C$_U$,  C$_B$, C$_V$ ,C$_R$ and C$_I$ are the color terms in the transformation equations , and X denotes the airmass.

On each test night, we got more than 40 frames data for each band. Each frame contains more than five standard stars on average, so for each band we have more than 200 records. Then, we derived the coefficients and fitted the observed magnitudes with {\it mknobsfile} and {\it fitparams} tasks in the {\it photcal} package in IRAF. The second-order terms of extinction were too small, so it has been ignored. we obtained transformation coefficients based on Equations from 3 to 7, including the zero points (Z), filter names, first-order extinction coefficients (K$^{'}$), filter names, root mean square (RMS) values and color terms (C), these parameters are listed in Table 6. We compare our calibration coefficients with results of Xinglong station in Table 7. One can see that some coefficients is smaller, may due to the altitude of Nanshan station is higher.

\begin{center}
\begin{tabular}{cccccc}
\multicolumn{6}{c}{{\bf Table 6} Coefficients and Standard deviation of NOWT photometric system}\\
\hline
Date & Filter & Zero points (Z) & Extinction ($K^{'}$) & Color Term (C) & RMS\\
\hline 
2017-5-03 & U$_{i}$ & 4.801 $\pm$ 0.039 & 0.496 $\pm$ 0.032 & -0.166 $\pm$ 0.010 & 0.132 \\
2017-5-04 & U$_{i}$ & 4.844 $\pm$ 0.036 & 0.540 $\pm$ 0.029 & -0.167 $\pm$ 0.009 & 0.126 \\
2017-8-29 & U$_{i}$ & 4.782 $\pm$ 0.015 & 0.550 $\pm$ 0.011 & -0.045 $\pm$ 0.009 & 0.072 \\
\hline
2017-5-03 & B$_{i}$ & 1.987 $\pm$ 0.015 & 0.256 $\pm$ 0.012 & -0.024 $\pm$ 0.006 & 0.061 \\
2017-5-04 & B$_{i}$ & 2.060 $\pm$ 0.019 & 0.264 $\pm$ 0.015 & -0.030 $\pm$ 0.006 & 0.062 \\
2017-8-29 & B$_{i}$ & 1.958 $\pm$ 0.007 & 0.359 $\pm$ 0.005 & -0.014 $\pm$ 0.003 & 0.042 \\
\hline
2017-5-03 & V$_{i}$ & 2.181 $\pm$ 0.013 & 0.166 $\pm$ 0.010 &  0.013 $\pm$ 0.005 & 0.047 \\
2017-5-04 & V$_{i}$ & 2.184 $\pm$ 0.012 & 0.216 $\pm$ 0.010 &  0.014 $\pm$ 0.004 & 0.040 \\
2017-8-29 & V$_{i}$ & 2.154 $\pm$ 0.006 & 0.258 $\pm$ 0.004 &  0.012 $\pm$ 0.002 & 0.029 \\
\hline
2017-5-03 & R$_{i}$ & 2.317 $\pm$ 0.015 & 0.104 $\pm$ 0.012 &  0.019 $\pm$ 0.009 & 0.053 \\
2017-5-04 & R$_{i}$ & 2.312 $\pm$ 0.015 & 0.156 $\pm$ 0.012 &  0.019 $\pm$ 0.008 & 0.050 \\
2017-8-29 & R$_{i}$ & 2.281 $\pm$ 0.007 & 0.192 $\pm$ 0.005 &  0.027 $\pm$ 0.005 & 0.034 \\
\hline
2017-5-03 & I$_{i}$ & 3.274 $\pm$ 0.012 & 0.078 $\pm$ 0.010 &  0.012 $\pm$ 0.004 & 0.047 \\
2017-5-04 & I$_{i}$ & 3.255 $\pm$ 0.013 & 0.128 $\pm$ 0.010 &  0.013 $\pm$ 0.004 & 0.044 \\
2017-8-29 & I$_{i}$ & 3.246 $\pm$ 0.005 & 0.143 $\pm$ 0.003 &  0.011 $\pm$ 0.002 & 0.023 \\
\hline
\label{table:coef}
\end{tabular}
\end{center}

\begin{center}
\begin{tabular}{ccccccc}
\multicolumn{7}{c}{{\bf Table 7} Extinction Coefficiets at Nanshan station }\\
\hline
Year        & K$^{'}_{U}$        & K$^{'}_{B}$        & K$^{'}_{V}$        & K$^{'}_{R}$        & K$^{'}_{I}$        & Ref\\
\hline
2016        &    0.590$\pm$0.022 & 0.431$\pm$0.029 & 0.282$\pm$0.026 & 0.217$\pm$0.019 & 0.156$\pm$0.021 & (\citealt{Bai:2018RAA})\\
2011-2012   &  & 0.348$\pm$0.022 & 0.236$\pm$0.017 & 0.168$\pm$0.019 & 0.085$\pm$0.021 &  (\citealt{Huang:2012RAA})\\
2008        &  & 0.330$\pm$0.007 & 0.242$\pm$0.005 & 0.195$\pm$0.004 & 0.066$\pm$0.003 &  (\citealt{ZhouAY:2009})\\
2006-2007   &  & 0.307$\pm$0.009 & 0.214$\pm$0.008 & 0.161$\pm$0.008 & 0.091$\pm$0.008 &  (\citealt{Huang:2012RAA})\\
2004-2005   &  & 0.296$\pm$0.012 & 0.199$\pm$0.009 & 0.141$\pm$0.010 & 0.083$\pm$0.009 &  (\citealt{Huang:2012RAA})\\
\hline
\label{table:coef_his}
\end{tabular}
\end{center}

In Figure 10, the comparison between the transformation equations in UBVRI bandsis and the Landolt standard magnitudes was shown .

\begin{figure}
    \begin{center}
    \includegraphics[width=15cm]{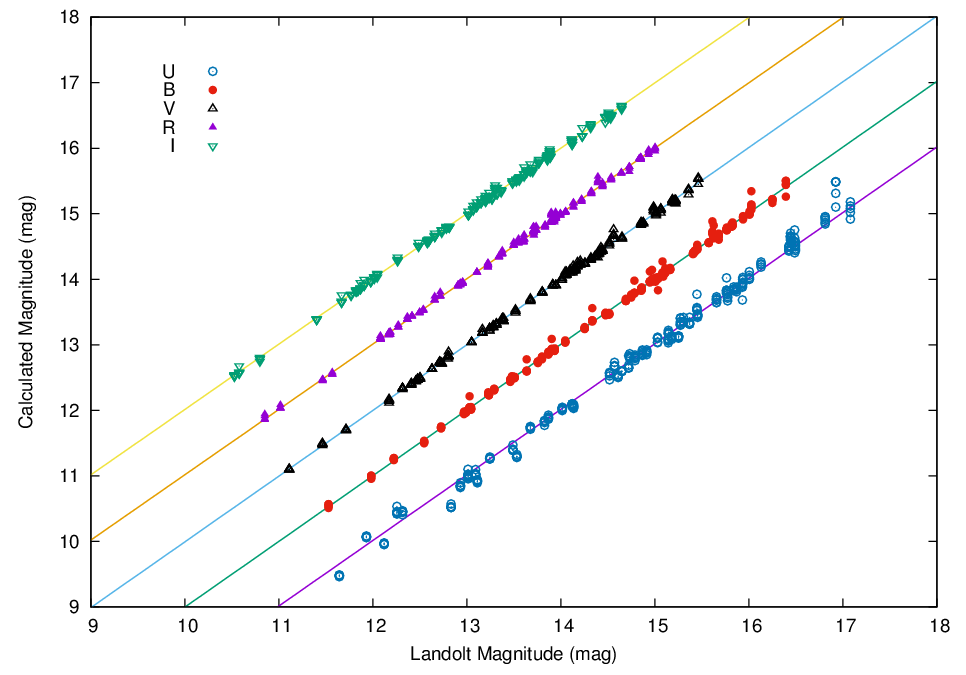}
    \caption[Fig. 4.1]{{\small Transformed magnitudes versus Landolt standards using derived transformation coefficients in five bands. The U, B, R, I bands transformed magnitudes are shifted to -2, -1, +1, +2 magnitudes. The best linear fits were represented with solid lines.}}
    \end{center}
    \label{fig:stand_coeff}  
\end{figure}

Using the transformation coefficients and the data which were taken on Aug. 2017 and May, the relationship between the Landolt standard and the shifted calibration  were derived. From Figure 10, we can see that the slopes of all fitted lines are close to 1.0. As listed in Table 6, the magnitude rms in each filter is smaller than 0.13. It indicate that the calibration between NOWT's photometric system and the Johnson system is established well.

\section{OVERALL PERFORMANCE}
\label{sect:PERFORMANCE}

\subsection{Throughput}

By observing several Landolt standards, entire optical system's total throughput can be estimated. Beside atmospheric transformation, quantum efficiency of camera and and filter response, the throughput also includes correctors and primary mirror reflection. The throughput of Lulin observatory was tested by the method (\citealt{Kinoshita:2005ChJAA}) which has been introduced by kinoshita. The NOWT's throughput efficiency is calculated based on the flux of Vega (\citealt{bohlin:2004}, \citealt{bohlin:2014}) that followed Equation (8), which introduced by Fan (\citealt{Fan:2016PASP}).

\begin{eqnarray}
\eta(\lambda) = \frac{F_{ADU}\cdot G}{F_{\lambda} \cdot \delta\lambda \cdot S_{tel}}
\end{eqnarray}

Where F$_{ADU}$ is the counts of standard star per second (ADU s$^{-1}$); from standard star's AB mag, we can derived its theoretical photon number F$_{\lambda}$ of per second (photon s$^{-1}$ cm$^{-2}$ {\AA}$^{-1}$); G represents the gain of camera (e$^{-}$ ADU$^{-1}$); $\delta\lambda$ is the grating dispersion of spectroscopic observations or the effective band of the filter in photometric observations({\AA}); $\lambda$ is the wavelength to computed the efficiency of spectroscopy or the effective wavelength for filters ({\AA}) and S$_{tel}$ is the primary mirror's effective area (cm$^{2}$).

 In Table 8, The median throughput of 5 bands are listed. Because our CCD chip is a blue sensitivity kind, the efficiency of blue side is higher than red side. We cleaned the main mirror before entering the winter, so the efficiency of Mar. 2018 is higher than May. 2017.

\begin{center}
\begin{tabular}{cccccc}
\multicolumn{6}{c}{{\bf Table 8} The total throughput of NOWT in UBVRI bands}\\
\hline
Date & U & B & V & R & I \\
\hline
2017-05-03 & 5.4\% & 25.8\% & 31.4\% & 25.4\% & 15.6\%\\
2018-03-10 & 6.1\% & 30.9\% & 38.6\% & 30.6\% & 18.2\%\\
\hline
\label{table:througt}
\end{tabular}
\end{center}

\subsection{ Photometry Accuracy And Limits}
The sky brightness at Nanshan station has been introduced by Hu (\citealt{Hu:2017ART}). Equation (9) is our way to make the estimation (\citealt{Howell:2000}).

\begin{eqnarray}
SNR &=& \frac{N_{star}}{\sqrt{N_{star}+n_{pix}(N_{sky}+N_{dark}+N_{readout}^{2})}} \\
\sigma &=& 1.0857 / SNR .
\end{eqnarray}

Where N$_{star}$ is the total photons of target, n$_{pix}$ is the pixels number under certain seeing condition, N$_{sky}$ is the sky background total number on each pixel, N$_{dark}$ is the dark current per pixel land, N$_{readout}$ is the readout noise listed in Section \ref{sec:gainRN}. Dark current's effect is extreme small, so it was neglected here. The photometric error can get easily from SNR, as $\sigma$ in Equation (10). The correction term between error and electron in magnitude is 1.0857.

\begin{figure}
    \begin{minipage}[t]{0.5\linewidth}
    \includegraphics[width=7cm]{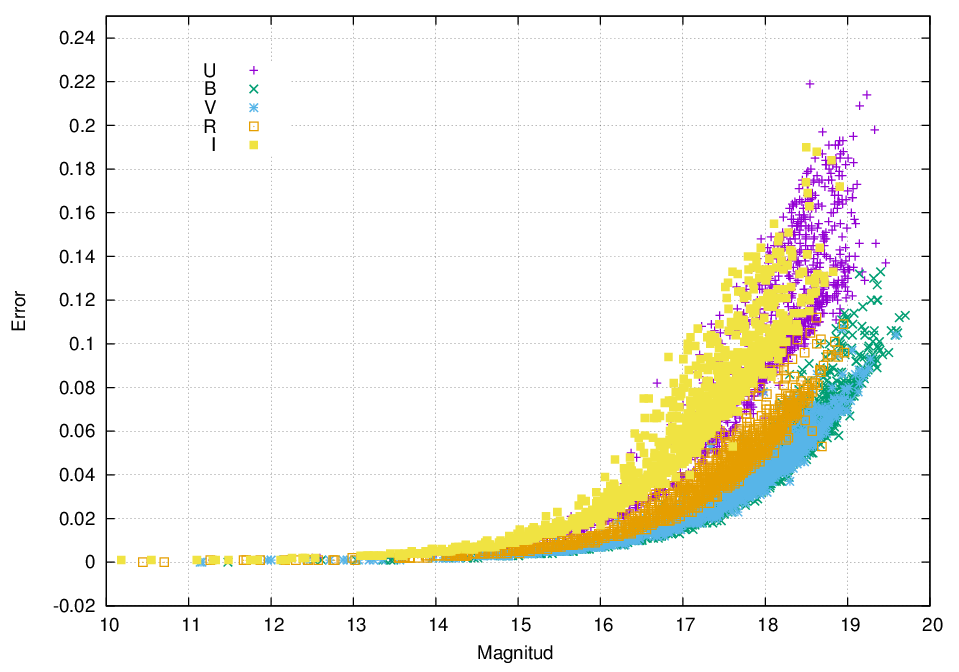}
    \caption[Fig. m13]{{\small Photometry error versus transformed magnitude of M13 with 600 s for U, 120 s for B, 144 s for V, 72 s for R and I band, respectively.}}
    \label{fig:ubvrim13}  
    \end{minipage}
    \begin{minipage}[t]{0.5\linewidth}
    \includegraphics[width=7cm]{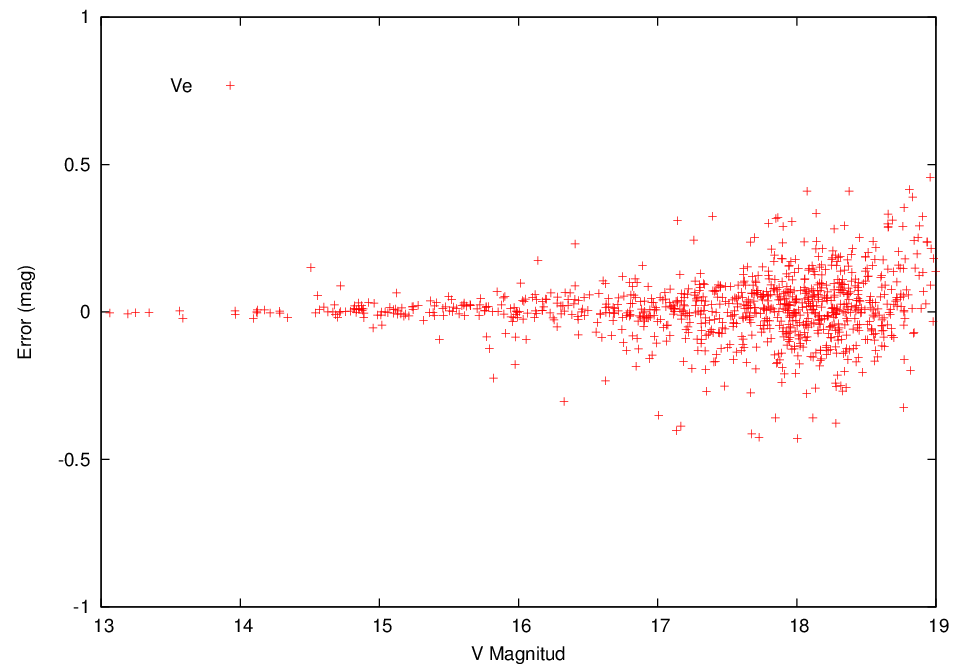}
    \caption[Fig. accuracy]{{\small Photometic accuracy errors of the two independent in the V band, the exposure time is 72 s.}}
    \end{minipage}
    \label{fig:accu}  
\end{figure}

To check the calibration relationship's reliability, observations were carried out by took targeting globular cluster M13 during several night in May 2017. In order to obtain the relationship between the exposure time, magnitude and signal-to-noise ratio, this part of the work is based on the IRAF standard photometric results. First, we took short exposure for each band, and then aligned and combined the image after preprocessing. The purpose is to improve the signal-to-noise ratio and avoid pixel saturation. The corresponding exposure time for each band was 600 s for U, 120 s for B, 144 s for V, and 72 s for R and I band. In Figure \ref{fig:ubvrim13}, the relationship between the photometry error of long time exposure and transformed stars brightness of different bands is shown. The photometry error of the I band corresponding to the magnitude is lower than other bands, and even lower than the U band. The main reason is that our CCD camera is blue sensitive.

After that, we set exposure time of M13 with V band  to 72 s, and took two images in a row, then photometry information of all stars was acquired, which SNR must be higher than 2. In Figure 12, the photometric accuracy \add(errors $\triangle$)V versus magnitude of the two independent measurements is plotted. The result is similar to the one meter telescope at Weihai Observatory (\citealt{Hu:2014RAA}) and Xinglong Station (\citealt{Bai:2018RAA}).

Different exposure time were made for each filter with five sets in the whole photometirc night (clear, breeze and no Moon). During the observations looped, the airmass of each star field was changing. After get the image and standard reduction, photometric accuracy and magnitude was obtained with different arimasses for each band. So, we can derive the relationship between the SNR of the limiting magnitudes changing with the airmass and exposure time. In Figure 13 and 14, the limiting magnitude versus the exposure time is plotted with SNR of 100 and 200. Observers can estimate their targets' exposure time by using these figures. Based on previous observations, we simulated the 300-second exposure when the telescope was aimed at the zenith. The simulate results with SNR of 5 are 19.0, 21.2, 21.3, 20.9 and 20.5 mag in the U, B, V, R and I bands, respectively.

\begin{figure}
    \begin{minipage}[t]{0.5\linewidth}
        \includegraphics[width=7cm]{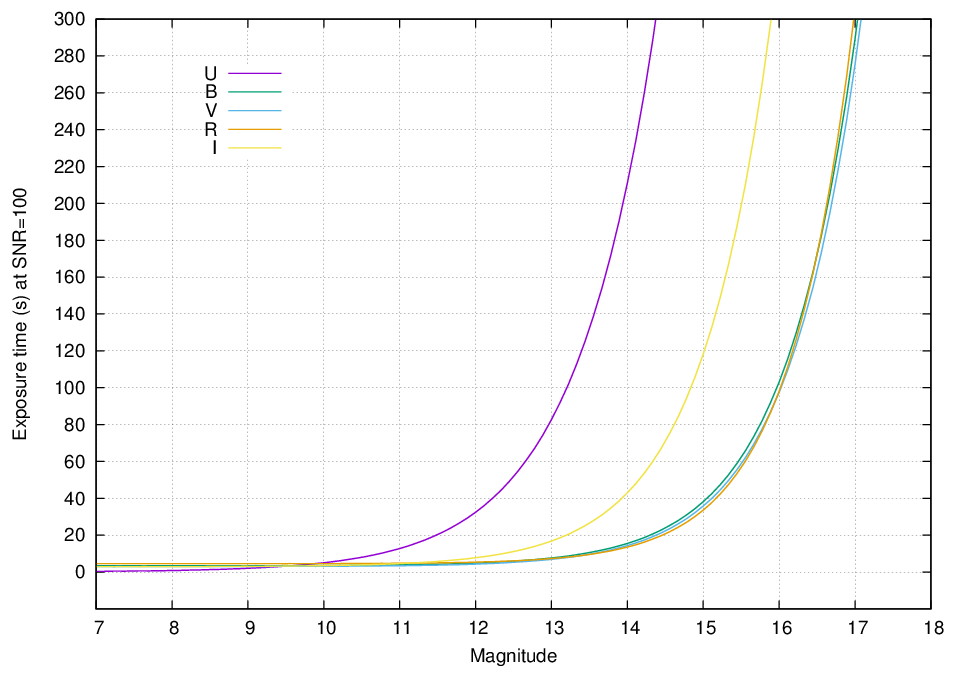}
        \caption[Fig. SN100]{{\small Exposure time requirement of limiting magnitude when SNR is 100.}}
    \label{fig:SN100}
    \end{minipage}
    \begin{minipage}[t]{0.5\linewidth}
        \includegraphics[width=7cm]{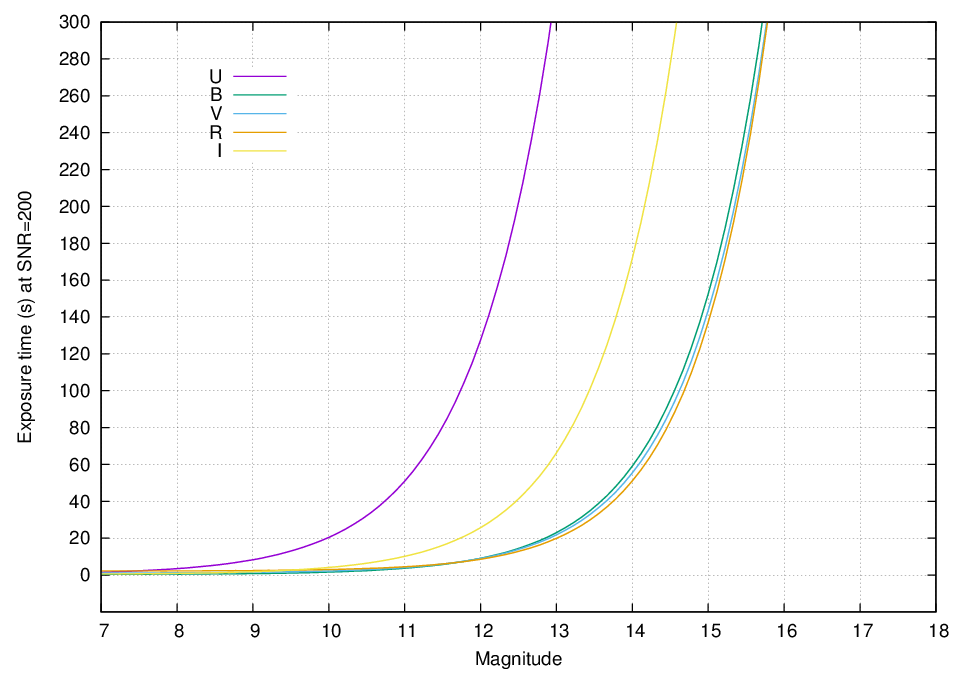}
        \caption[Fig. SN100]{{\small Exposure time requirement of limiting magnitude when SNR is 200.}}
    \end{minipage}
    \label{fig:SN200}
\end{figure}

\section{SUMMARY}
\label{sect:SUMMARY}

The NOWT's photometric system is introduced in this paper. In order to offer more valuable information for users, the readout noise, bias, dark current, gain and linearity of the CCD camera is tested. One can get that the camera has a very good scientific-grade CCD from the test results. We selected dozens of Landolt standards for multiple night observations in different seasons. Based on thees test data, the transformation coefficients are derived between the standard magnitude and the instrumental UBVRI magnitude. The color terms and atmospheric extinction coefficients are found and compared the result with other station. Due to the blue sensitivity chip and the altitude of Nanshan station, the NOWT's throughput is good, especially on the blue side. The limiting magnitude can reach 19.0, 21.2, 21.3 , 20.9, 20.5 mag in corresponding 5 bands with the exposure of 300s and SNR of 5. At last, we give guidance regarding exposure to observers about the limiting magnitude at different exposure time and SNRs requirement.

\begin{acknowledgements}
We gratefully acknowledge the support of the staff of the NOWT. This research is supported by the program of the light in China's Western Region (LCWR; grant Nos. 2015-XBQN-B-04, 2015-XBQN-A-02), the National Natural Science Foundation of China (grant Nos. 11803076, 11873081, 11661161016 and U1831209), the 13th Five-year Informatization Plan of Chinese Academy of Sciences (grant No. XXH13503-03-107), the Youth Innovation Promotion Association CAS (grant Nos. 2014050, 2018080), and the Strategic Priority Research Program of the Chinese Academy of Sciences (grant No. XDB23040100), 2017 Heaven Lake Hundred-Talent Program of Xinjiang Uygur Autonomous Region of China  and the Open Project Program of the Key Laboratory of Optical Astronomy(NAOC).
\end{acknowledgements}

\bibliographystyle{raa}
\bibliography{R2}

\end{spacing}

\end{document}